\documentstyle[twoside,fleqn,espcrc2,epsfig]{article}

\newcommand{\AmS}{{\protect\the\textfont2
  A\kern-.1667em\lower.5ex\hbox{M}\kern-.125emS}}
\def\kreiso{\lower0.85pt\hbox{\Large $\bullet$}}
\def\kreisv{\raise0.85pt\hbox{$\scriptstyle\bigcirc$}}
\def\bbox{\lower0.85pt\hbox{$\Box$}}
\newcommand{\blacksquare}{\multiput(0,0)(0.25,0){22}{\line(0,1){6}}\;\:\,}

\title{\vspace{-2.65cm}
       {\normalsize DESY 97--178}    \\[-0.2cm]
       {\normalsize September 1997}   \\
       \vspace{1.25cm}
       Lattice Chiral Schwinger Model in the Continuum 
       Formulation\thanks{Talk given by V. Bornyakov}}

\author{V. Bornyakov\address{Institute for High Energy Physics,
        RU-142284 Protvino, Russia},
        G. Schierholz\address{Deutsches Elektronen-Synchrotron DESY, 
        D-22603 Hamburg and D-15735 Zeuthen, Germany} 
        and
        A. Thimm\address{Institut f\"ur Theoretische Physik, Freie
        Universit\"at Berlin, D-14195 Berlin, Germany}}
       
\begin{document}

\begin{abstract}
We pursue further an approach to lattice chiral fermions in which
the fermions are treated in the continuum. To render the
effective action gauge invariant, counterterms have to be introduced.
We determine the counterterms for smooth gauge fields, both
analytically and numerically. The final result is that the imaginary
part of the effective action can be computed analytically from the 
lattice gauge field, while the real part is given by one half of the
action of the corresponding vector model. 
\end{abstract}

\maketitle

\section{INTRODUCTION}

In formulating chiral gauge theories on the lattice, it has been 
suggested~\cite{GS1} to discretize only the gauge fields and treat the 
fermions in the continuum. One starts
from a lattice with spacing $a$. This is the lattice on which the
simulations will be done. Then one constructs a finer lattice with
spacing $a_f$. On this lattice one puts the fermions. Before one can 
do this, one has to extrapolate the gauge fields to the interior of the
original lattice. This was done in~\cite{GKSW}. The method makes use
of Wilson fermions to remove the doublers. 
One then computes the effective fermionic action in the limit 
$a_f \rightarrow 0$, while keeping $a$ fixed. This action
will in general not be invariant under chiral gauge transformations,
but it will already be close. So close that chiral gauge
invariance can be restored by simply adding a few local
counterterms to the action. For similar ideas see~\cite{Bodwin}.

In this talk we shall restrict ourselves to gauge fields with zero
topological charge. We start from the fermionic action
\begin{eqnarray*}
\lefteqn{S_{\pm} = \frac{1}{2a}\sum_{x,\mu} \left\{\bar{\psi}(x) \gamma_{\mu}
 [(1+P_{\pm}U_{\mu}(x))\psi(x+\mu)\right. } \\
& & \left.- (1+P_{\pm}U^{\dagger}_{\mu}(x-\mu))\psi(x-\mu)]\right\} + S_W,
\end{eqnarray*}
\begin{eqnarray*}
\lefteqn{S_W = \frac{1}{2a}\sum_{x,\mu} \bar{\psi}(x) 
[2\psi(x)-U_{\mu}(x)\psi(x+\mu)} \\
& &  - U^{\dagger}_{\mu}(x-\mu) \psi(x-\mu)] 
\end{eqnarray*}
with $P_\pm = (1 \pm \gamma_5)/2$. Later on we will also consider an
ungauged Wilson term, $S_W$, with $U_\mu \equiv 1$. The effective action is
given by
\begin{eqnarray*}
\exp(-W_\pm) = \int {\cal D}\bar{\psi}{\cal D}\psi \exp(-S_\pm).
\end{eqnarray*}

Let us denote the anomaly free effective action generically by the subscript 
$\scriptsize a$.
We are now looking for an action of the form
\begin{eqnarray*}
W^\Sigma_\pm = W_{\pm} + \mbox{counterterms},
\end{eqnarray*}
so that
\begin{eqnarray*}
\widehat{W}_a = \lim_{a_f \rightarrow 0} W^\Sigma_a
\end{eqnarray*}
is invariant under chiral gauge transformations.
It will then turn out that 
\begin{eqnarray*}
\mbox{Re} \widehat{W}_\pm = \frac{1}{2} (W + W_0), 
\;\mbox{Im} \widehat{W}_a
= \lim_{a_f \rightarrow 0} \mbox{Im} W_a, 
\end{eqnarray*}
where $W$ is the effective action of the corresponding vector model,
and $W_0$ is the free action, both taken at $a_f \rightarrow 0$. The
imaginary part of $W_\pm$ has been
computed analytically~\cite{GS2}. It depends only on the zero gauge
field modes (torons) of the background field (see below).

\section{EFFECTIVE ACTION}

For the extrapolation of the gauge fields we use~\cite{GKSW}. The
effective action is computed by means of the Lanczos method. The
Lanczos vectors are re-orthogonalized after every iteration. 

The gauge field in its most general form can be written 
\begin{equation}
 A_{\mu}(x) = \frac{2\pi}{L} t_{\mu} + \varepsilon_{\mu\nu} \partial_{\nu}
 \alpha(x) + \mathrm{i} g^{-1}(x) \partial_{\mu} g(x),
\label{field}
\end{equation}
where it is assumed that $A_\mu(x) \in [-\pi,\pi)$, and where 
$t_{\mu}$ are the zero momentum modes (torons), 
$\partial^2 \alpha(x) = F_{12}(x)$ and $g(x) \in U(1)$ is a gauge
transformation. 
We assume periodic boundary conditions
for the gauge fields and antiperiodic
boundary conditions for the fermions.

\subsection*{\it Toron Field}
\vspace*{0.3cm}

Let us first consider the case 
\begin{eqnarray*}
 A_{\mu}(x) = \frac{2\pi}{L} t_{\mu} 
+ \mathrm{i} g^{-1}(x) \partial_{\mu} g(x), 
\end{eqnarray*}
where $g(x)$ is a small gauge transformation. Under a small gauge
transformation we understand a transformation that does not
change $t_\mu$. The effective action $\lim_{a_f \rightarrow 0} W_\pm$
is not gauge invariant. It is easy to identify the appropriate
counterterm. It is
\begin{equation} 
c \sum_{x} A_\mu^2(x). 
\label{counter}
\end{equation}
The coefficient $c$ can be computed analytically. We find $c = -
0.0202$. We will use this counterterm throughout the paper.
In Fig.~1 we plot the real and imaginary part of $W^\Sigma_{-}$
as a function of $(a_f/a)^2$ for a particular toron field and $g(x) =
0$. We do not expect anomalous contributions in this case, so it is
legitimate to consider one species of fermions only. These results are
compared with $(W+W_0)/2$ and the analytical result for
$\mbox{Im}\widehat{W}_{-}$~\cite{GS2}. We see that the real part 
converges rapidly to $(W+W_0)/2$, while the imaginary
part is practically equal to its analytic value for all $a_f$. We have
also considered an ungauged Wilson term. In this case the imaginary part
converges less rapidly to its continuum value. We find that real and
imaginary part of the effective action are gauge invariant in the
limit $a_f \rightarrow 0$. We will show a picture in the next section.

\begin{figure}[thb]
\begin{centering}
\epsfig{figure=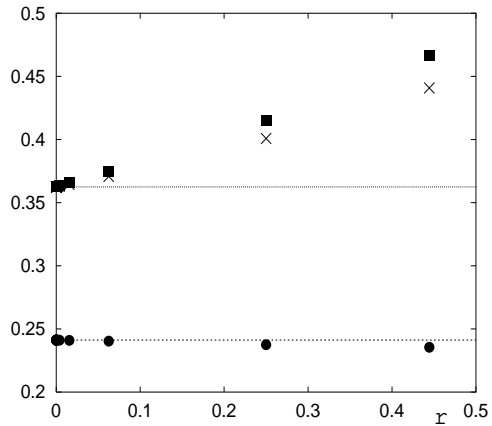,height=5.8cm,width=7.0cm}
\vspace{-0.7cm}
\caption{$\mbox{Re}W^\Sigma_{-}-W_0$ (\,\lower0.85pt\hbox{$\blacksquare$}),
$\mbox{Im}W_{-}$ ($\kreiso$) as well as $(W-W_0)/2$ ($\times$) as a function
of $r=(a_f/a)^2$. The lines are the continuum 
results.}
\vspace{-0.50cm}
\end{centering}
\end{figure}

\subsection*{\it Fluctuating Fields}
\vspace*{0.3cm}

Next we consider a general gauge field as given by
(\ref{field}) with $F_{12} \neq 0$. This configuration has been
generated by a Monte Carlo method at $\beta = 6.0$ on a lattice small
enough to avoid singular plaquettes. The average plaquette value at
this coupling was $\approx 0.9$. With the counterterm (\ref{counter})
we find similar results as in Fig.~1. In particular, we find
that $\mbox{Im}\widehat{W}_a$ 
is in complete agreement with the analytic
result, meaning that the imaginary part depends alone on the
magnitude of the toron field $t_\mu$.

To test for gauge invariance we applied small random gauge transformations
to the gauge field. To monitor the variation of the effective action
under such gauge transformations we introduce the measure
\begin{eqnarray*} 
\Delta X = \frac{1}{N} \sum_{\{g\}} |X^g - X|,\; X = \mbox{Re}W_\pm,
\mbox{Im}W_{\pm,a}, 
\end{eqnarray*}
where the sum is over a set of $N$ gauge transformations, $X$ is
the initial result, and $X^g$ is the result after the gauge
transformation $g$. The effect of these gauge transformations is shown
in Fig.~2. We see that the real part of $W^\Sigma_{-}$ becomes
gauge invariant in the limit $a_f \rightarrow 0$. For the imaginary
part we have to distinguish between the anomalous and the anomaly free
model. In the 
anomalous case the imaginary part is not gauge \hfill invari-

\newpage
\begin{figure}[thb]
\begin{centering}
\epsfig{figure=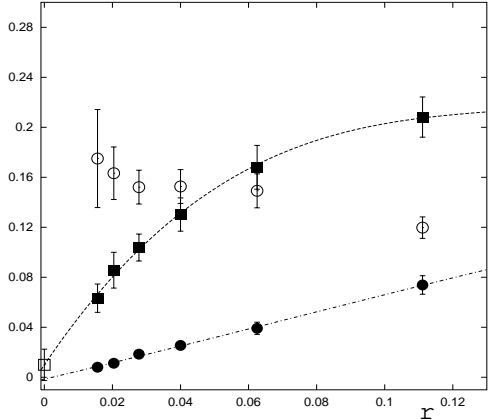,height=5.8cm,width=7.0cm}
\vspace{-0.7cm}
\caption{$\Delta\mbox{Im}W_{-}$ ($\kreisv$), 
$\Delta\mbox{Im}W_a$ ($\kreiso$) and $\Delta(\mbox{Re}W^\Sigma_{-}-W_0)$
(\,\lower0.85pt\hbox{$\blacksquare$}) as a function
of $r=(a_f/a)^2$. The lines are polynomial fits.}
\vspace{-0.70cm}
\end{centering}
\end{figure}

\noindent
ant, and that was also not expected. For the anomaly free model
we take $e_{-} = (1,1,1,1)$ and $e_{+} = 2$, $e_\pm$ being the charges
of the right and left handed fermions, respectively. In this case we
find that the imaginary part of the effective action is gauge
invariant in the limit $a_f \rightarrow 0$.

\subsection*{\it Toron Field in Singular Gauge}
\vspace*{0.3cm}

Let us now go back to the toron field and allow for a gauge transformation
which transforms $A_\mu(x)$ to
\begin{eqnarray*}
 A_{\mu}(x) = 2 \pi \delta_{x_\mu, 1} t_\mu\;[\mbox{mod}\;2\pi].
\end{eqnarray*}
The reason for considering such a transformation was to test our
result for the imaginary part of the effective action under different
conditions. Suppose that $|t_\mu| > 1/2$. For charge 2 it is sufficient
that $|t_\mu| > 1/4$. For the anomalous model we then find that 
the analytic result changes under this transformation, unlike in the
previous case, due to the compactness of the gauge field. Our lattice
results show exactly this behavior. The real part of the effective action
was found to be gauge invariant.

\subsection*{\it Vortex-Antivortex Configuration}
\vspace*{0.3cm}

Another large gauge transformation which changes $t_{\mu}$ is 
$g(x)=\exp(\mbox{i}h(x))$ with
\newpage
\begin{eqnarray}
\lefteqn{h(x) = 2\pi [\frac{x_2-v_2+1}{L}+\frac{1}{2}] 
 [\theta(x_1-v_1-1)} \nonumber \\
& & -\theta(x_1-\bar{v}_1-2)] - 2\pi [\frac{x_1-\bar{v}_1-1}{L}+\frac{1}{2}]
\label{gauge}  \\
& & \times [\theta(x_2-v_2)-\theta(x_2-\bar{v}_2-1)], \; [\mbox{mod}\; 2\pi]. 
\nonumber
\end{eqnarray}
This transformation creates a vortex-antivortex pair at $x = v$ and
$\bar{v}$, respectively. With the counterterm (\ref{counter}) it
turns out that under this gauge
transformation neither the real part, nor the imaginary part of the
effective action are invariant in the limit $a_f \rightarrow 0$. This
holds for the anomalous as well as for the anomaly free model. The 
good news is however that the
imaginary part agrees with the analytic result which changes by
exactly the same amount. As far as the real part is concerned, this
indicates that further counterterms containing derivatives of the
gauge field are needed to restore gauge invariance. It should be noted
that this problem does not exist in the case of non-compact gauge
field action because there (\ref{gauge}) is not a gauge transformation.

\section{CONCLUSIONS}

For smooth gauge fields we have found an action which is ready to use
in numerical simulations. The real part of the effective action can be
expressed in terms of the action of the corresponding vector model,
while the imaginary part can be computed analytically from the lattice
gauge field.

\end{document}